\documentclass[preprint,review,12pt,sort&compress]{elsarticle}
\usepackage[margin=1.18 in]{geometry}
\usepackage{multirow}
\usepackage[usenames, dvipsnames]{color}
\definecolor{UW}{RGB}{64, 38, 96}

\usepackage{amssymb}

\usepackage{float}

\journal{Composite Structures}

\begin{document}

\begin{frontmatter}


\cortext[cor1]{Corresponding Author, \ead{g-cusatis@northwestern.edu}}

\title{Mode I and II Interlaminar Fracture in Laminated Composites: A Size Effect Study}


\author[address1]{Marco Salviato}
\author[address2]{Kedar Kirane}
\author[address3]{Zden\v ek P. Ba\v zant}
\author[address3]{Gianluca Cusatis\corref{cor1}}

\address[address1]{William E. Boeing Department of Aeronautics and Astronautics, University of Washington, Seattle, Washington 98195, USA}
\address[address2]{Department of Mechanical Engineering, Stony Brook University, Stony Brook, NY 11794, USA}
\address[address3]{Department of Civil and Environmental Engineering, Northwestern University, Evanston, IL 60208, USA}

\begin{abstract}
\linespread{1}\selectfont

This work investigates the mode I and II interlaminar fracturing behavior of laminated composites and the related size effects. Fracture tests on geometrically scaled Double Cantilever Beam (DCB) and End Notch Flexure (ENF) specimens were conducted to understand the nonlinear effects of the cohesive stresses in the Fracture Process Zone (FPZ). 
The results show a significant difference between the mode I and mode II fracturing behaviors. It is shown that, while the strength of the DCB specimens scales according to the Linear Elastic Fracture Mechanics (LEFM), this is not the case for the ENF specimens. 
Small specimens exhibit a pronounced pseudo-ductility with limited size effect and a significant deviation from LEFM, whereas larger specimens behave in a more brittle way, with the size effect on nominal strength closer to that predicted by LEFM. This behavior, due to the significant size of the Fracture Process Zone (FPZ) compared to the specimen size, needs to be taken into serious consideration. It is shown that, for the specimen sizes investigated in this work, neglecting the non-linear effects of the FPZ can lead to an underestimation of the fracture energy by as much as $55\%$, with an error decreasing for increasing specimen sizes. Both the mode I and II test data can be captured very accurately by Ba\v zant's type II Size Effect Law (SEL). 

\end{abstract}


\end{frontmatter}


\section{Introduction}
\label{intro}

Laminated polymer matrix composites are extensively used across the main engineering fields, from automotive, aerospace and civil engineering to microelectronics \cite{automotive,aerospace,civil,stenzen,pascault,takahashi,blok,shan,plasticsbook,Ceccato}. This is thanks to their excellent in-plane mechanical properties, which make composites the material of choice for the manufacturing of large lightweight structures \cite{aerospace, stenzen, pascault, takahashi}. However, broad implementation of laminated composites is limited by their pathologically low macro-scale delamination resistance, which can trigger other damage mechanisms and lead to structural collapse. 

Although several approaches such as the use of three-dimensional woven and braided reinforcements \cite{Cox, Long, Bogdanovich1, Bogdanovich2, Pankow}, fiber stitching \cite{Mouritz} or architected adhesives \cite{Fleck} have been proposed, throughout the years, to mitigate this problem, the interlamianar fracture resistance remains a weakness of polymer composites. Therefore, the design of large composite structures necessitates prediction of the critical loads for the onset of interlaminar fracturing and the related structural scaling laws. This scaling, often overlooked in the literature, is the key toe determine the material fracture properties governing delamination and to correlate the mechanical behavior in small-scale laboratory tests to the delamination resistance of large composite structures. 

This scaling, however, cannot be captured by conventional theories such as plasticity or Linear Elastic Fracture Mechanics (LEFM). In fact, due to the emergence of several micro-damage mechanisms such as the sub-critical crack formation in the matrix, crack deflection and fiber-tow bridging, the size of the Fracture Process Zone (FPZ) in front of a stress-free, interlaminar crack is often not negligible. Accordingly, the interlaminar fracturing behavior and the associated energetic size effect on structure strength can be described only if a certain length scale is linked to the finite size of the nonlinear FPZ and is used alongside the proper values of the interlaminar fracture energy and strength. Unfortunately, estimating these material properties using standard testing procedures is quite challenging since the crack onset is often followed by snap-back instability or discontinuous crack propagation.

A possible way to overcome these issues is leveraging size effect testing \cite{bazant1998_1,salviato2016_1}. This study presents an experimental and numerical investigation of the efficacy of the mode I and II size effect testing to characterize the interlaminar fracturing behavior of composites. It is shown that the size effect characterization enables an accurate estimation of the interlaminar fracture energy and the size of the FPZ. Furthermore, since the size effect analysis requires measuring only the peak loads, no visual inspection of the crack tip location is needed. Not only does this make size effect testing easier and more accurate than other methods, but it also allows overcoming the problems with snap-back instabilities typically afflicting mode II delamination tests.  

\section{Materials and Methods}

\subsection{Materials and preparation}
\label{materialprep}

The experiments were conducted on woven composite specimens manufactured by compression molding. A DGEBA-based epoxy resin was chosen as polymer matrix whereas the reinforcement was provided by a twill 2x2 fabric made of carbon fibers. The main in-plane mechanical properties were characterized by testing [$0^{\circ}$]$_8$ and [$45^{\circ}$]$_8$ coupons under uniaxial tension following ASTM standard procedures \cite{ASTMD3039, ASTMD3518}. The results of this characterization are described in detail in \cite{salviato2016_1, Sal2016b, Sal2016c, Kirane2016a, Kirane2016b} and are summarized in Table \ref{T1}.

\subsection{Specimen characteristics}

To investigate the interlaminar fracturing behavior of laminated composites and the related scaling, size effect tests were conducted on geometrically scaled specimens. Mode I Interlaminar fracture tests were conducted on Double Cantilever Beam (DCB) specimens whereas the Mode II interlaminar fracturing behavior was investigated using  
End Notch Flexure (ENF) specimens. For all the loading configurations, geometrically-scaled specimens of three different sizes were tested. The layup was maintained as $\left[0^{\circ}\right]_n$ for all the tests, with $n$ varying as a function of the desired thickness of the specimen. For all the specimens, the initial crack was obtained by inserting a $10$-$\mu$m-thick teflon film during the lamination process. 

\begin{figure}[H]
\center
\includegraphics[trim= 0cm 0cm 0cm 0cm,width=16cm]{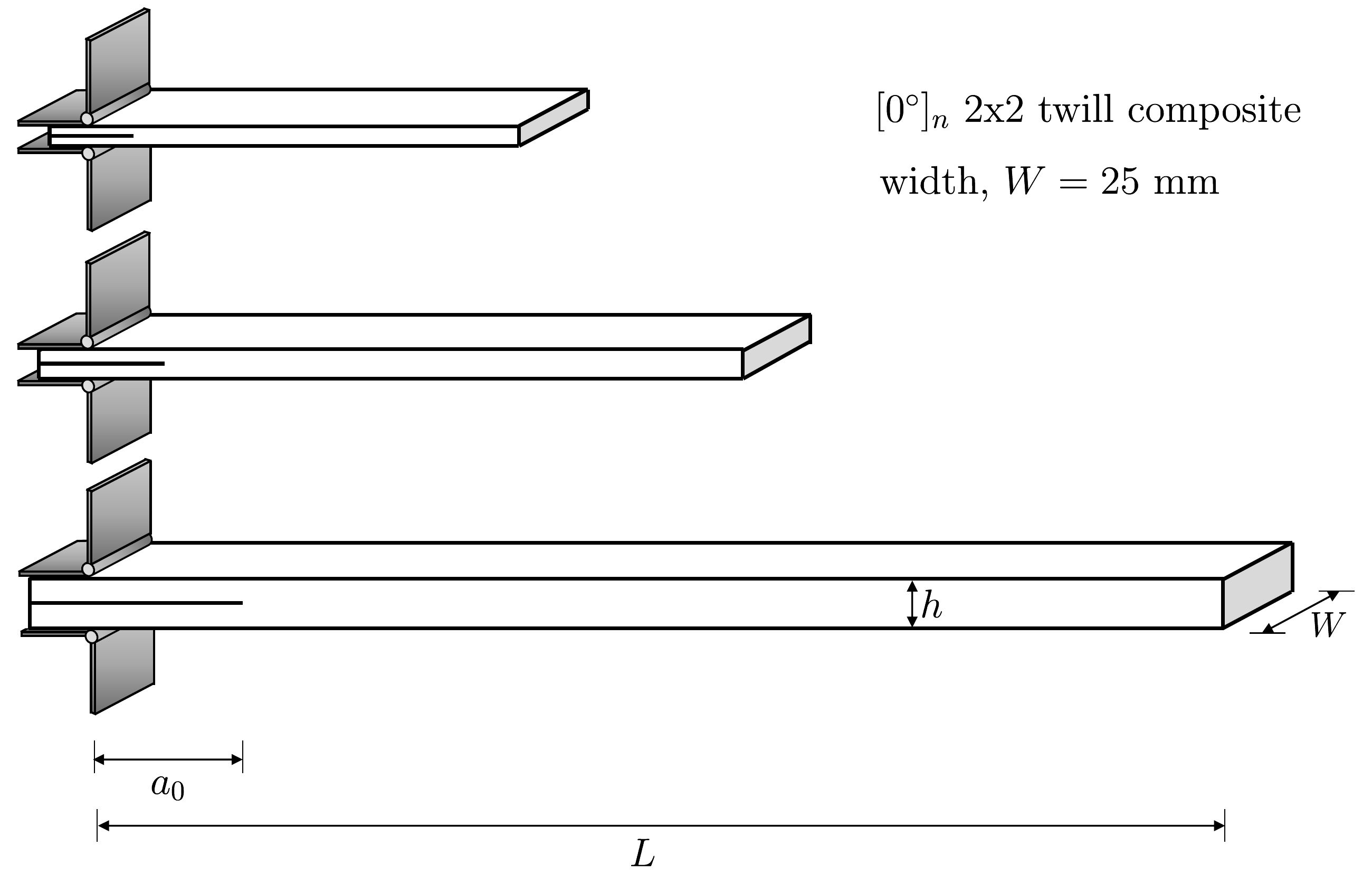}
\caption{Geometry of the Double Cantilever Beam (DCB) specimens under study.}
\label{DCB_geometry}
\end{figure}

As schematically represented in Figure \ref{DCB_geometry}, the smallest DCB specimen investigated in this work was made of eight layers, $\left[0^{\circ}\right]_8$, which resulted in total thickness $h=1.9$ mm. Based on ASTMD5528 \cite{ASTMD5528}, the gauge length of the specimen was $L=100$ mm while the width was $W=25$ mm and the initial interlaminar crack length was $a_0=40$ mm. For the other specimens, all the geometrical features were geometrically-scaled as $1:3:5$, except for the width, $W$, which was kept constant and equal to $25$ mm (see Figure \ref{DCB_geometry}). The specimen characteristics are summarized in Table \ref{T2}. 

\begin{figure}[H]
\center
\includegraphics[trim= 0cm 0cm 0cm 0cm,width=16cm]{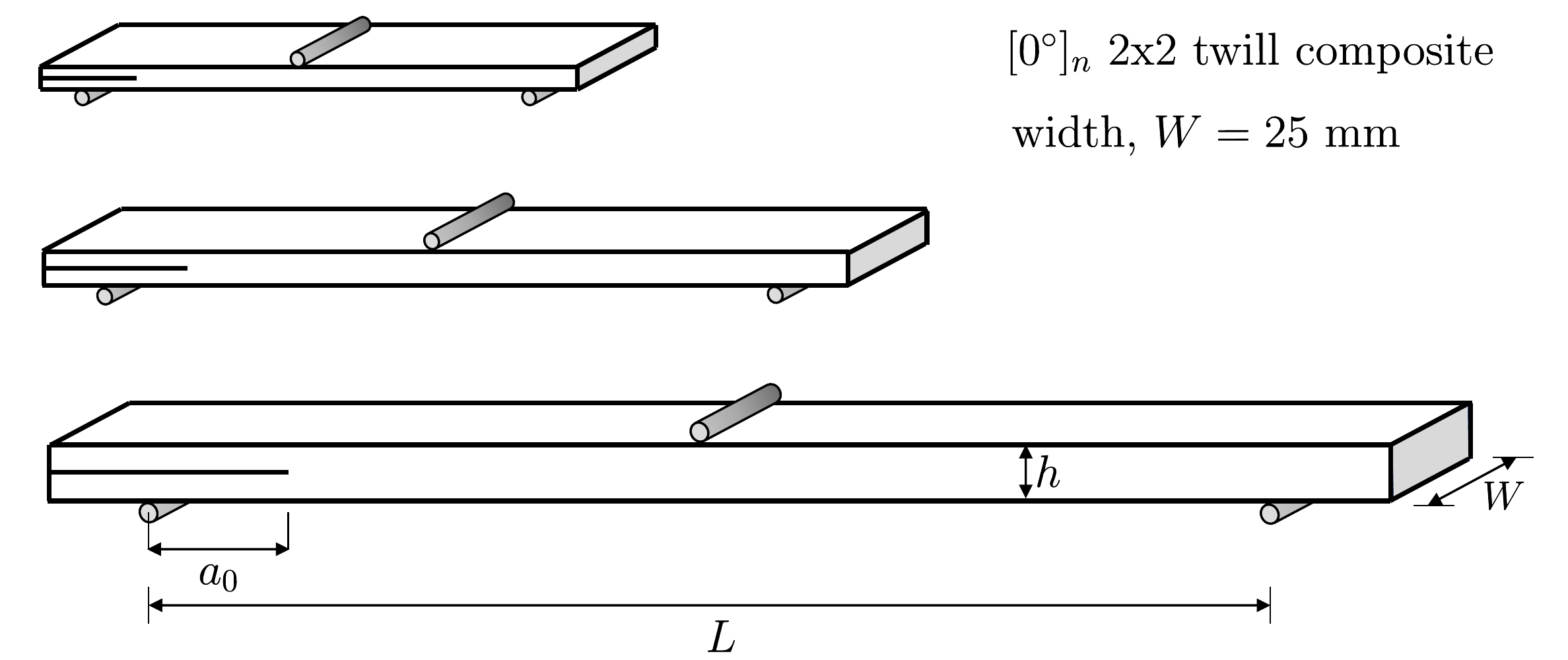}
\caption{Geometry of the End Notch Flexure (ENF) specimens under study.}
\label{ENF_geometry}
\end{figure}

Same as in \cite[e.g.]{Cantwell97,Sousa15}, End Notch Flexure (ENF) specimens were used to characterize the mode II interlaminar fracturing behavior. The smallest ENF specimens featured a gauge of length $L=200$ mm, and an initial interlaminar crack of length $a_0=50$ mm. For the other specimens, all the geometrical features were geometrically-scaled as $2:3:5$, except for the width, $W$, which was kept constant and equal to $25$ mm (see Figure \ref{ENF_geometry}). The specimen characteristics are summarized in Table \ref{T3}. 

\subsection{Testing}

The DCB and ENF specimens were tested in a closed-loop servohydraulic MTS machine with a $5$ kN load-cell. The tests were conducted at a constant crosshead rate (stroke control, $0.5$mm/min for the smallest specimens), the rate being adjusted for the different sizes to achieve roughly the same strain rate. With such settings, the test lasted no longer than approximately $10$ min for all the specimens. Stroke, force, and loading time were recorded with a sampling frequency of $10$ Hz. A DIC system from Correlated Solutions \cite{correlated} consisting of a $5$ MP digital camera and a workstation for image postprocessing was used to measure the displacement field in the specimen with an acquisition frequency of $1$ Hz.

\section{Experimental Results}

\subsection{Mode I interlaminar fracture tests}

\begin{figure}[H]
\center
\includegraphics[trim= 0cm 0cm 0cm 0cm,width=\textwidth]{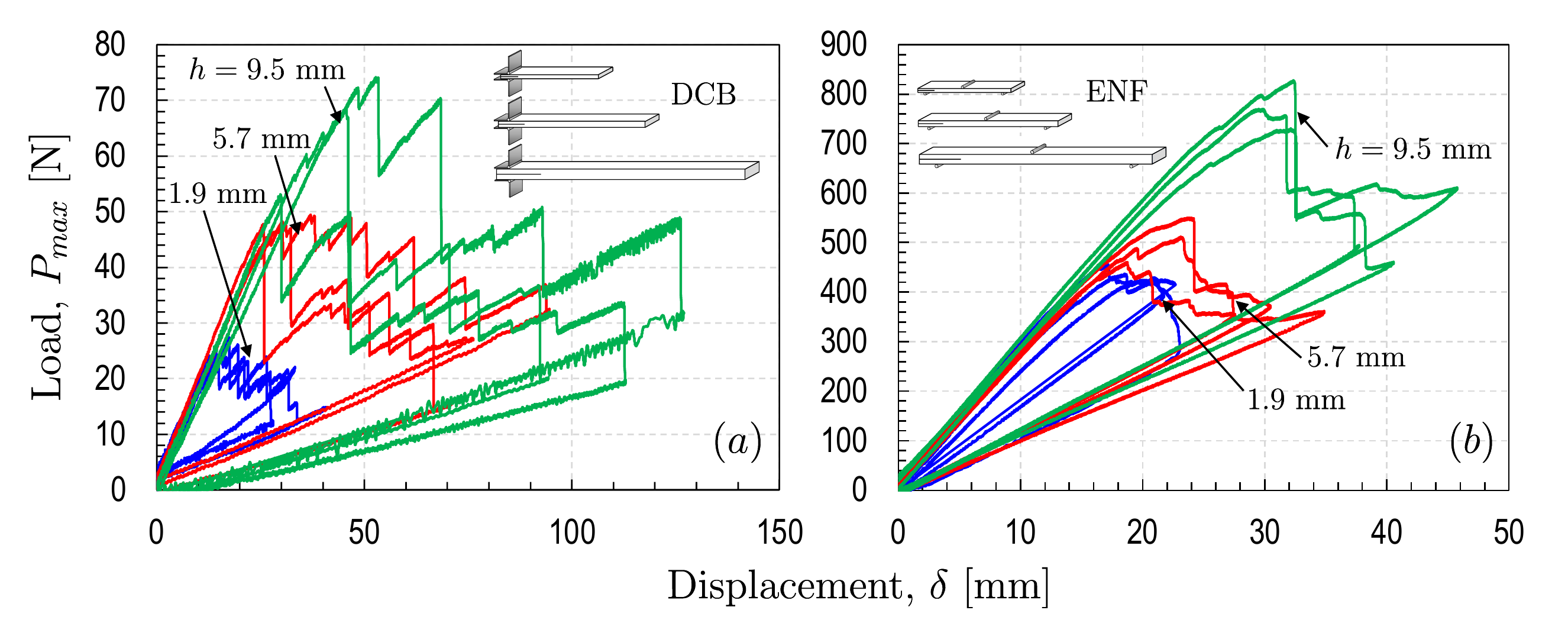}
\caption{Typical load-displacement curves of geometrically-scaled showing decreasing nonlinearity increasing specimen dimensions: (a) DCB, (b) ENF specimens. }
\label{load_displ}
\end{figure}

After the completion of the mode I fracture experiments, the load and displacement data were analyzed. Figure \ref{load_displ}a shows, for the various investigated sizes, the typical load-displacement curves, whereas the peak loads and related structural strengths calculated as $\sigma_{Nc}=P_{max}/hW$ are summarized in Table \ref{T4}. As can be noted, for all the specimen sizes investigated in this work, the structural behavior before reaching the maximum load is almost linear with very limited damage preempting the subsequent crack propagation. This is an indication of pronounced brittleness and of a limited effect of the nonlinear cohesive stresses in the Fracture Process Zone (FPZ) on the structural behavior. 

It is interesting that, probably due to the waviness of the tows of the twill $2\times 2$ fabric, the crack propagation occurred in several unstable jumps. This phenomenon was clearly detected during the tests using Digital Image Correlation (DIC) analysis of the displacement field surrounding the crack tip, and is manifested in Figure \ref{load_displ}a by distinct load drops and recoveries after reaching the maximum load. This phenomenon, not uncommon in the delamination testing of textile composites, made the detection of the exact crack tip location very cumbersome (which is also true for, e.g., concrete). As a matter of fact, the ASTMD5528 standard \cite{ASTMD5528} does not recommend the study of crack propagation by visual techniques in the presence of such behavior. 

However, as it will be clear in the following, studying the fracturing behavior by Size Effect, requires only the knowledge of the peak load. There is no need to locate the crack tip at any time. This makes the proposed methodology easier and accurate even in situations in which the visual inspection of the crack tip location is not recommended.

\subsection{Mode II interlaminar fracture tests}

\begin{figure}[H]
\center
\includegraphics[trim= 0cm 0cm 0cm 0cm,width=\textwidth]{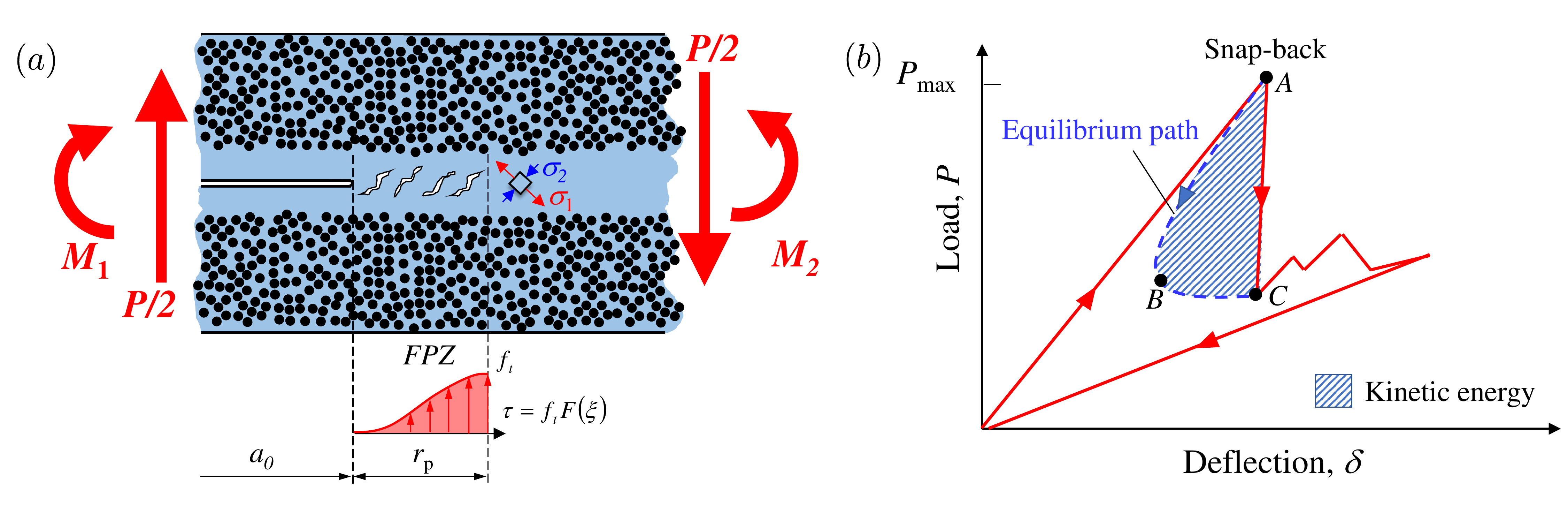}
\caption{(a) Schematic representation of the damage mechanisms in the FPZ of a mode II interlaminar crack leading to emergence of nonlinear cohesive shear stresses; (b) schematic illustration of the snap-back instability affecting the ENF tests.}
\label{micro}
\end{figure}

The load-displacement curves obtained from the ENF tests are represented in Figure \ref{load_displ}b for all the specimen sizes while Table \ref{T5} summarizes the peak loads and structural strengths, $\sigma_{Nc}=P_{max}/hW$, reported in the tests. It is interesting to note that, in contrast to the DCB specimens, the load-displacement curves of the ENF specimens exhibit a more pronounced nonlinear behavior before the peak load. This phenomenon, which is more significant for small specimen sizes, is related to the formation of a nonlinear FPZ whose size is not negligible compared to the structure size. For sufficiently small specimens, the nonlinear damage in the FPZ in the form of sub-critical, matrix microcraking \cite{Carraro1}, crack deflection and fiber pullout can affect the structural behavior significantly. A typical process of formation of an FPZ under mode II loading conditions in thermoset polymers is schematically represented in Figure \ref{micro}a. 

It is worth noting from Figure \ref{load_displ}b that the mode II interlaminar crack propagated unstably right after the peak load was reached for all the investigated sizes. This unstable crack propagation is associated with a snap-back instability, as indicated by the sudden drop of the load after the peak. It should be highlighted here that the foregoing snap-back instability makes the calibration of cohesive laws for mode II delamination particularly challenging. In fact, since the load frame cannot follow the equilibrium load path (curve ABC in Figure \ref{micro}b), the measured load-displacement curve exhibits a sudden dynamic  drop schematized by the vertical segment AC in Figure \ref{micro}b. Accordingly, the shaded area ABC in Figure \ref{micro}b represents the kinetic energy induced to the system by the load frame. Calibrating the cohesive law such that the experimental curve be matched exactly cannot lead to an accurate characterization of the cohesive behavior since it only allows estimating an upper bound for the initial fracture energy. On the other hand, leveraging only the data after the snap-back instability to infer the cohesive behavior can only lead to the calibration of the parameters describing the last part of the cohesive law, which control the formation of the fully developed FPZ.

The size effect tests presented in this work allow overcoming the foregoing issues since they only require the characterization of the peak load for geometrically-scaled specimens of different sizes. Thanks to this procedure, the initial mode II fracture energy can be easily estimated using the equations presented in the next sections. Further, this information can be used to estimate the initial part of the cohesive curve precisely, using the approach outlined by Cusatis and Schauffert in \cite{Cusatis1}.

\section{Analysis and Discussion}

In polymer composites the size of the non-linear Fracture Process Zone (FPZ) occurring in the presence of a large stress-free, interlaminar crack is generally not negligible. The stress field along the FPZ is nonuniform and decreases with crack opening, due to discontinuous cracking, micro-crack deflection, micro-crack pinning or fiber/tow bridging of the crack \cite{Salviatoc,Carraro1}. As a consequence, the fracturing behavior and, most importantly, the energetic size effect associated with the given structural geometry, cannot be described by means of classical Linear Elastic Fracture Mechanics (LEFM). To capture the effects of a finite, non-negligible FPZ, the introduction of a characteristic (finite) length scale, related to the fracture energy and the strength of the material, is necessary \cite{Baz84,Baz90,bazant1998_1,bazant1996_1,salviato2016_1,Cusatis1,Yao2017a,Yao2018b,Yao2018c,Salviatoc,Carraro1}. This is done in the following sections.

\subsection{Size effect law for mode I and II interlaminar fracturing in composites}

In quasibrittle composites the effects of the nonlinear FPZ on the interlaminar fracturing behavior can be analyzed leveraging an equivalent linear elastic fracture mechanics approach. To this end, an effective crack length $a=a_0+c_f$ with $a_0=$ initial crack length and $c_f=$ effective FPZ length is considered. Following LEFM, the energy release rate can be written as follows:
\begin{equation}
G^{(i)}\left(\alpha\right)=\frac{\sigma_N^2h}{E^*}g_{(i)}(\alpha)
\label{eq:Gf}
\end{equation}
where $\alpha=a/h=$ normalized effective crack length, $\sigma_N=P/Wh=$ nominal stress, $E^*= $ equivalent elastic modulus, and $g_{(i)}\left(\alpha\right)=$ dimensionless energy release rate for mode-$i$. The failure condition can now be written as:
\begin{equation}
G^{(i)}\left(\alpha_0+c_f^{(i)}/h\right)=\frac{\sigma_{Nc}^2h}{E^*}g_{(i)}\left(\alpha_0+c_f^{(i)}/h\right)=G_f^{(i)}
\label{failure}
\end{equation}
where $G_f^{(i)}$ with $i=I,II$ is the mode-$i$ fracture energy of the material and $c_f^{(i)}$ is the effective FPZ length, assumed to be a material property. It should be remarked that this equation characterizes the peak load conditions if $g_{(i)}'(\alpha)>0$, i.e., only if the structure has positive geometry \cite{bazant1998_1} (which means that [$\partial G(a) / \partial a]_P > 0$).

By approximating $g_{(i)}\left(\alpha\right)$ with its Taylor series expansion at $\alpha_0$ and retaining only up to the linear term of the expansion, one obtains:
\begin{equation}
G_f^{(i)}=\frac{\sigma_{Nc}^2h}{E^*} \left[g_{(i)}(\alpha_0)+\frac{c_f^{(i)}}{h}g_{(i)}'(\alpha_0)\right]
\label{Taylor}
\end{equation}
which can be rearranged as follows \cite{bazant1998_1}:
\begin{equation}\label{SEL}
\sigma_{Nc}=\sqrt{\frac{E^*G_f^{(i)}}{hg_{(i)}(\alpha_0)+c_f^{(i)}g_{(i)}'(\alpha_0)}}
\label{Sel}
\end{equation}
Here $g_{(i)}'\left(\alpha_0\right)=\mbox{d}g_{(i)}\left(\alpha_0\right)/\mbox{d}\alpha$.

This equation relates the nominal strength of radially scaled structures to a characteristic size, $h$ and it can be rewritten in the following form:
\begin{equation}
\sigma_{Nc}=\frac{\sigma_{0}^{(i)}}{\sqrt{1+h/h_0^{(i)}}}
\label{eq:sigmaNc2}
\end{equation}
where $\sigma_0^{(i)}=\sqrt{E^*G_f^{(i)}/c_f^{(i)}g_{(i)}'(\alpha_0)}$ and $h_0^{(i)}=c_f^{(i)}g_{(i)}'(\alpha_0)/g_{(i)}(\alpha_0)=$ constant, depending on both FPZ size and specimen geometry. Contrary to classical LEFM, Eq. (\ref{eq:sigmaNc2}) is endowed with a characteristic length scale $h_0^{(i)}$. This is the key to describe the transition from pseudo-ductile to brittle behavior with increasing structure size reported especially in the mode II fracture tests.

\subsubsection{Fitting of the experimental data by SEL}

The values of $G_f^{(i)}$ and $c_f^{(i)}$ can be determined by regression analysis of the experimental data. Following Ba\v{z}ant \emph{et al}. \cite{bazant1998_1} the following transformation is used:
\begin{equation}\label{linear}
X=h,\quad Y=\sigma_{Nc}^{-2}
\end{equation}
\begin{equation}\label{param}
\sigma_0^{(i)}=\left[C^{(i)}\right] ^{-1/2},\quad h_0^{(i)}=\frac{C^{(i)}}{A^{(i)}}=\frac{1}{A^{(i)}\left[\sigma_0^{(i)}\right]^2}
\end{equation}
thanks to which Eq. (\ref{eq:sigmaNc2}) can now be expressed in the following linear form:
\begin{equation}
Y=A^{(i)}X+C^{(i)}
\label{eq:slope1}
\end{equation}

Eq. (\ref{eq:slope1}) can be used to perform a linear regression analysis of the size effect data provided that all the specimens are scaled exactly so that $g_{(i)}(\alpha_0)$ and $g_{(i)}'(\alpha_0)$ take the same values for all the tests.

The parameters of the size effect law, $A^{(i)}$ and $C^{(i)}$, can be directly related to the mode-$i$ fracture energy of the material, $G_f^{(i)}$ and the effective FPZ size, $c_f^{(i)}$ as follows:
\begin{equation}
G_f^{(i)}=\frac{g_{(i)}\left(\alpha_0\right)}{E^*A^{(i)}},\quad c_f^{(i)}=\frac{C^{(i)}}{A^{(i)}}\frac{g_{(i)}\left(\alpha_0\right)}{g_{(i)}'\left(\alpha_0\right)}
\label{eq:GFCF}
\end{equation}
provided that the functions $g_{(i)}\left(\alpha\right)$ and $g_{(i)}'\left(\alpha\right)=\mbox{d}g_{(i)}\left(\alpha\right)/\mbox{d}\alpha$ and the elastic modulus $E^*$ are known. The calculation of $g_{(i)}\left(\alpha\right)$ and $g_{(i)}'\left(\alpha\right)$ is discussed in the next section.

\subsection{Calculation of $g_{(i)}(\alpha)$ and $g_{(i)}'(\alpha)$} \label{sec:gandgprimecalculation}

\begin{figure}[H]
\center
\includegraphics[trim= 0cm 0cm 0cm 0cm,width=16cm]{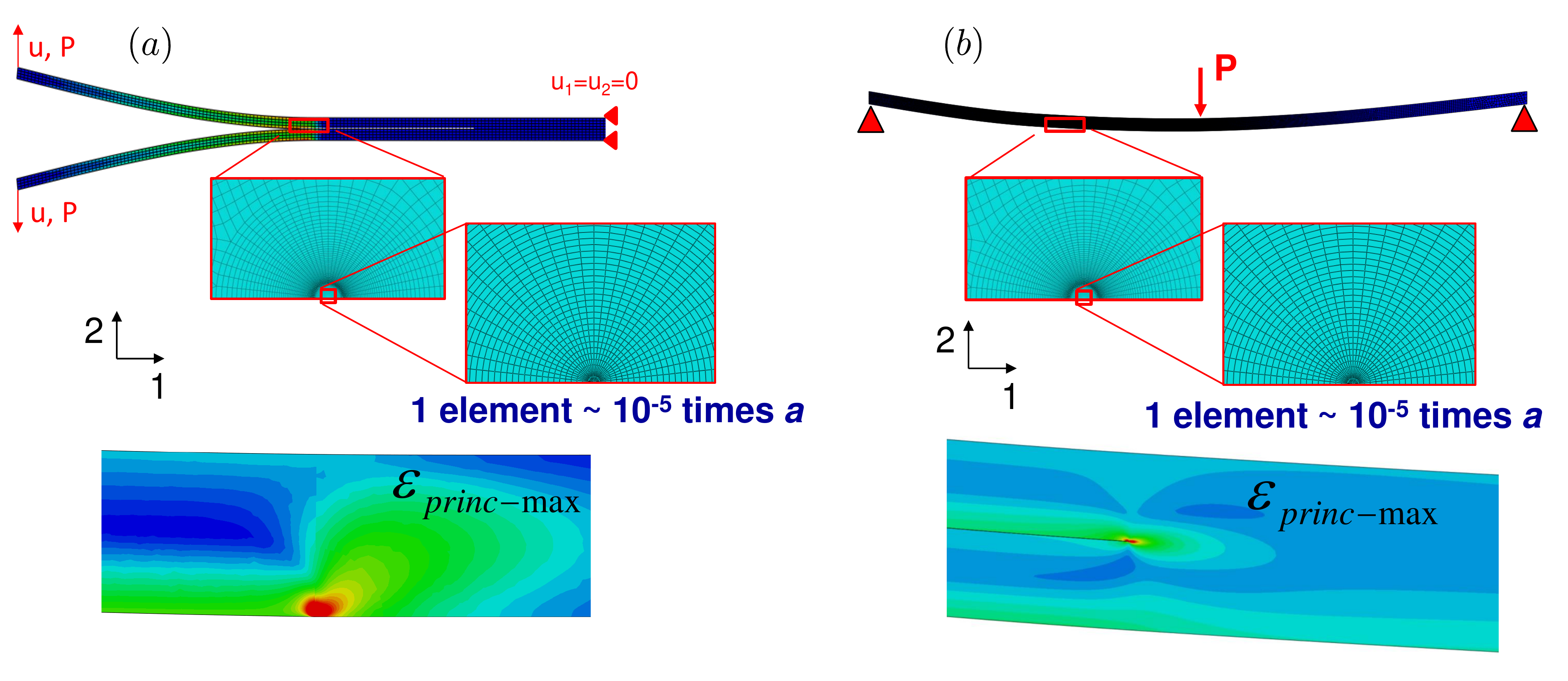}
\caption{Schematic representation of the Linear Finite Element Model for the calculation of $g\left(\alpha\right)$ and $g'\left(\alpha\right)$ and typical maximum principal strain fields: (a) DBC and (b) ENF specimens.}
\label{g_calculations}
\end{figure}

The function $g_{(i)}(\alpha)$ can be obtained through Finite Element (FE) analyses. In this work the simulations were performed in ABAQUS Implicit 6.13 \cite{abaqus} using 8-node biquadratic plain strain quadrilateral elements (CPS8), combined with the quarter element technique \cite{barsoum} at the crack tip to provide accurate results. The smallest element size at the tip was about $a_{0}\cdot10^{-5}$ leading to roughly 330,000 elements for the whole model. A linear elastic, orthotropic constitutive model was used for the simulations, with $E_1=53500$ MPa, $E_2=10000$ MPa and $\upsilon_{12}=0.3$ where, as shown in Figures \ref{g_calculations}a,b, directions $1$ and $2$ were longitudinal and orthogonal to the axis of the specimen, respectively. The J-integral approach \cite{rice1968_1} was used to estimate the energy release rate for both the DCB and ENF specimens. 

A representation of the meshes used in this work and of a typical contour plot of the maximum principal strain close to the crack tip is provided in Figures \ref{g_calculations}a,b. It may be mentioned that, for the simulation of the ENF specimens, the general contact formulation with penalty stiffness available in ABAQUS/implicit \cite{abaqus} was used to capture the frictional phenomena occurring between the crack surfaces. A friction coefficient $\mu=0.3$ was used although a comprehensive parametric study revealed that the energy dissipated by friction does not have a significant effect on the calculation of the energy release rate. In fact, the normal component of the forces acting on the crack surfaces in the incipient failure condition were found to be negligible. 

Finally, it should also be noted that the FE simulations allowed accounting explicitly for the rotation of the arms of the DBC specimens at the crack tip without the need of correction factors or approximations \cite{Williams1,Williams2}.

Once the J-integral was calculated from ABAQUS, the value of $g_{(i)}(\alpha)$ was obtained using the following expression based on LEFM:
\begin{equation}
g_{(i)}(\alpha)=\frac{G^{(i)}(\alpha)E^*}{h\sigma_N^2}
\label{eq:galpha}
\end{equation}
where $\sigma_{N}=P/Wh$, $P=$ applied load, $W=$ width, and $\alpha=a/h$ is the normalized effective crack length. For all the calculations of the dimensionless energy release functions, the condition $E^*=E_1$ was used.

To determine the function $g_{(i)}'(\alpha)$, various normalized crack lengths close to the selected value of $\alpha$ were considered in order to calculate the tangent slope of $g_{(i)}(\alpha)$ through linear interpolation \cite{salviato2016_1}. 

Following the foregoing procedure, the values of the dimensionless energy release rate functions for the mode-I interlaminar crack were found to be $g_{(I)}\left(\alpha_0\right)=45019.9$ and $g_{(I)}'\left(\alpha_0\right)=4125.8$ with $\alpha_0=a_0/h=21.05$. For the ENF specimens, on the other hand, $g_{(II)}\left(\alpha_0\right)=807.0$ and $g_{(II)}'\left(\alpha_0\right)=120.4$ with $\alpha_0=a_0/h=13.16$. 

\subsection{Analysis of the mode I interlaminar fracture tests by the Size Effect Law (SEL)}

\begin{figure}[H]
\center
\includegraphics[trim= 0cm 0cm 0cm 0cm,width=\textwidth]{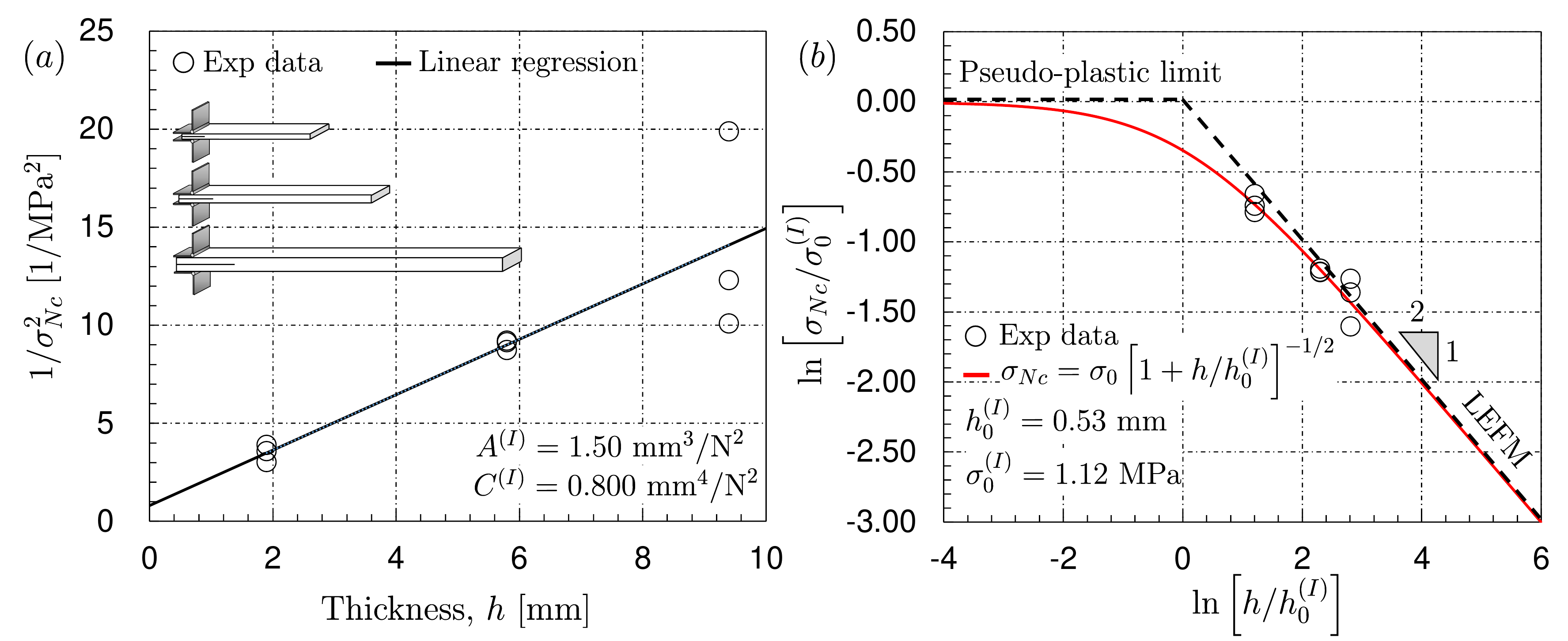}
\caption{Size effect study. (a) Linear regression analysis to characterize the size effect parameters. (b) Size Effect plot in Mode-I interlaminar fracture. }
\label{SEL_I}
\end{figure}

The mode I interlaminar fracture tests were analyzed by means of the type II size effect law outlined in the foregoing sections. Figure \ref{SEL_I}a shows the linear regression analysis based on the transformation reported in Eq. (\ref{linear}). This led to the identification of the following parameters: $A^{(I)}=1.50$ mm$^3$/N$^2$ and $C^{(I)}=0.800$ mm$^4$/N$^2$. As can be noted, the SEL is in very good agreement with the experimental data, notwithstanding the scatter of the data on the large size specimens. 

The results of the linear regression analysis can be used along with Eqs. (\ref{param}a,b) to estimate the values of the pseudo-plastic limit under mode I loading, $\sigma_0^{(I)}$, and the transitional thickness, $h_0^{(I)}$, marking transition from pseudo-ductile to brittle structural behavior. Based on the tests conducted in this work it was found that $\sigma_0^{(I)}=1.12$ MPa and $h_0^{(I)}=0.53$ mm. Using Eqs. (\ref{eq:GFCF}a,b), the estimated mode I fracture energy and related equivalent FPZ length are $G_f^{(I)}=0.56$ N/mm and $c_f^{(I)}=5.82$ mm respectively.

The best fit of the experimental data by the Size Effect Law, Eqs. (\ref{SEL}) or (\ref{eq:sigmaNc2}), is also represented in Figure \ref{SEL_I}b where the normalized strength, $\sigma_{NC}/\sigma_0^{(I)}$ is plotted as a function of the normalized characteristic thickness $h/h_0^{(I)}$ in double logarithmic scale. As can be noted from the figure, the SEL exhibits two important asymptotes. The horizontal asymptote represents the structural strength predicted by plastic limit analysis, for which the failure of geometrically similar structures should always occur when $\sigma_{NC} = \sigma_0^{(I)}$ and thus implies no scaling. In contrast, the asymptote of inclination $-1/2$ (in log-log scale) represents the structural strength predicted by the Linear Elastic Fracture Mechanics (LEFM) according to which the strength scales with $h^{-1/2}$. The intersection of the two asymptotes corresponds to the transitional thickness $h_0^{(I)}$ which marks the transition from pseudo-plastic to brittle behavior for increasing specimen sizes. 

It is important to note that the structural strength of all the DCB specimens investigated follows the LEFM asymptote relatively closely, meaning that the nonlinear cohesive stresses in the FPZ do not affect significantly the structural behavior for the specimens of the sizes investigated in this work or larger. This is also confirmed by the relatively good agreement between the mode I fracture energy value estimated by the SEL and its value estimated by the LEFM, which is $G_{f,LEFM}^{(I)} = \sigma_{Nc}^2h /E^*g_{(I)} \left(\alpha_0\right)$. The LEFM predicts  fracture energy values $G_{f,LEFM}^{(I)}=$ 0.46, 0.54 and 0.55 N/mm for the small, medium and large sizes,  respectively. Although the values estimated by the LEFM are slightly size dependent, the difference compared to SEL is always within the scatter of the experimental data, especially for the medium and large size specimens. 

The foregoing results confirm that, as suggested by the ASTMD5528 \cite{ASTMD5528}, LEFM can be used to accurately predict the mode I delamination onset of the twill $2\times 2$ specimens investigated in this work, as well as the related scaling.

The nonlinear effects of the FPZ would start to be significant only for ultra-thin composite structures \cite{Pellegrino1, Pellegrino2} whose thickness is typically lower than the transitional one: $h<h_0^{(I)}=0.53$ mm. 

\subsection{Analysis of the mode II interlaminar fracture tests by the Size Effect Law (SEL)}

\begin{figure}[H]
\center
\includegraphics[trim= 0cm 0cm 0cm 0cm,width=\textwidth]{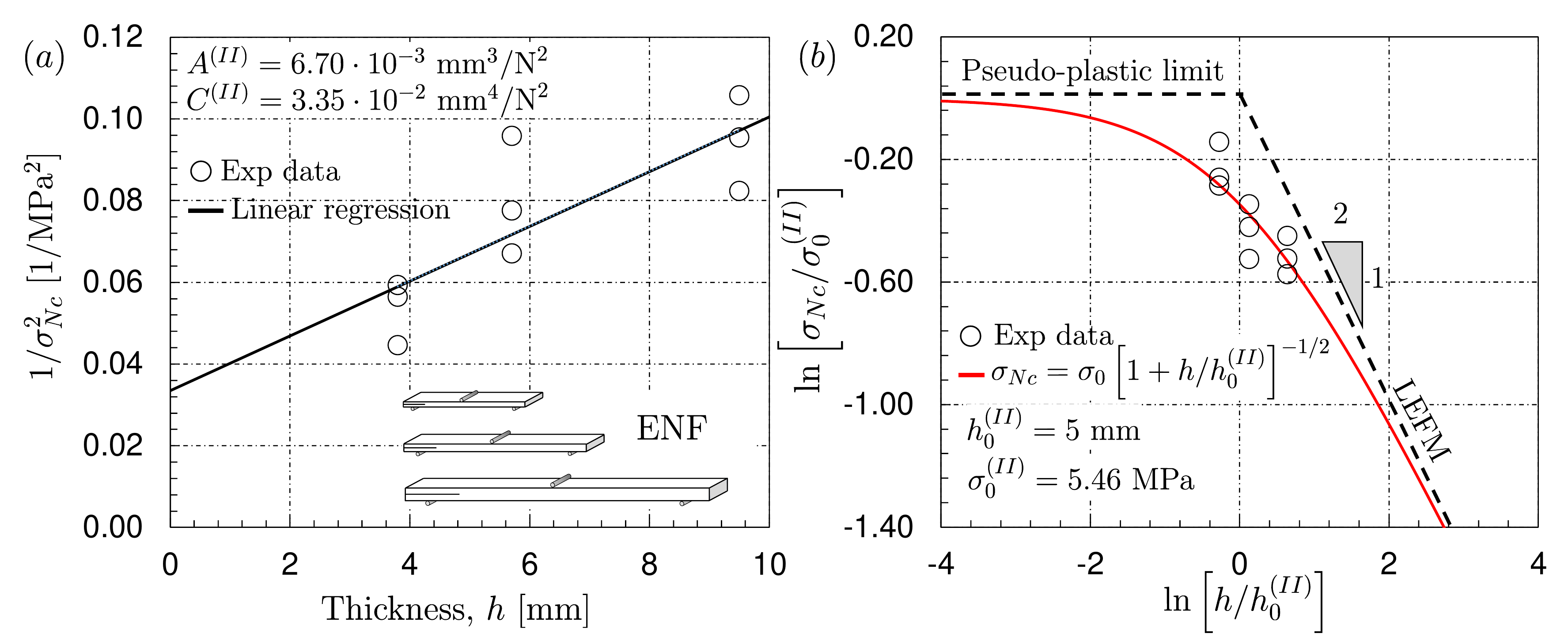}
\caption{Size effect study. (a) Linear regression analysis to characterize the size effect parameters. (b) Size Effect plot in Mode-II interlaminar fracture. }
\label{SEL_II}
\end{figure}

The mode II SEL formulated in the foregoing sections was used to analyze the tests on the ENF specimens, following a similar procedure as the one described for the analysis of the DCB specimens. Figure \ref{SEL_II}a shows the linear regression analysis leveraging the transformation reported in Eq. (\ref{linear}), with $A^{(II)}=6.70\cdot 10^{-3}$ mm$^3$/N$^2$ and $C^{(II)}=3.35\cdot 10^{-2}$ mm$^4$/N$^2$. As can be noted, also in this case the SEL is in excellent agreement with the experimental data, and the fitting enables estimating of the pseudo-plastic limit under mode II loading, $\sigma_0^{(II)}$, and the transitional thickness, $h_0^{(II)}$ by means of Eqs. (\ref{param}a,b). Based on the tests conducted in this work, it is found that $\sigma_0^{(II)}=5.46$ MPa and $h_0^{(II)}=5.00$ mm. On the other hand, the mode II fracture energy and related equivalent FPZ length are $G_f^{(II)}=2.25$ N/mm and $c_f^{(II)}=33.51$ mm, respectively.

The best fit of the experimental data by the Size Effect Law, Eqs. (\ref{SEL}) or (\ref{eq:sigmaNc2}), is represented in Figure \ref{SEL_II}b where the normalized strength, $\sigma_{NC}/\sigma_0^{(II)}$ is plotted against the normalized characteristic thickness $h/h_0^{(II)}$ in double logarithmic scale. It is interesting to note that, in contrast to the results on the DCB specimens, the structural strength values of the ENF specimens lie right at the transition between the pseudo-plastic and the LEFM asymptotes. 

This observation confirms that the structural behavior of the ENF specimens is much more influenced by the strain redistribution caused by the damage in the FPZ, and that the resulting structural behavior is much more pseudo-ductile. Thanks to the presence of a characteristic length scale, $c_f^{(II)}$, associated with the size of the Fracture Process Zone (FPZ), the SEL can capture the experimental data with excellent accuracy, as shown in Figure \ref{SEL_II}b. The same cannot be said of the LEFM nor the plasticity theory, which both lack of a characteristic length scale and thus cannot capture the transition from pseudo-ductile to brittle behavior, as demonstrated in the ENF tests. 

It is interesting to note that the use of LEFM to estimate the mode II fracture energy would lead to size dependent values, $G_{f,LEFM}^{(II)}$ being equal to $1.02$, $1.28$, and $1.36$ N/mm for the small, medium and large size specimens respectively. This is a serious issue since it contradicts the fundamental assumption of LEFM that $G_f^{(II)}$ are material properties. More importantly, the values of the fracture energy predicted by the LEFM are $54.9\%$, $43.0\%$ and $39.8\%$ lower than the value provided by the SEL depending on the specimen size. 

\subsection{FE analysis of the experiments via a Cohesive Zone Model (CZM)}

To validate the Mode I and II fracture energies estimated by SEL, a finite element analysis was conducted. The same mesh and boundary conditions as described in Section \ref{sec:gandgprimecalculation} were used whereas the cohesive stresses on the crack surfaces were captured by the standard cohesive interaction algorithm available in Abaqus/implicit \cite{abaqus}. Thanks to the fine mesh, the cohesive stresses in the FPZ were always described by at least $50$ elements, for any size, and for both the DCB and ENF specimens. In the absence of additional information on the exact shape of the cohesive curve, the traction-separation law was supposed to be linear for both mode I and II loading. 

\begin{figure}[H]
\center
\includegraphics[trim= 0cm 0cm 0cm 0cm,width=\textwidth]{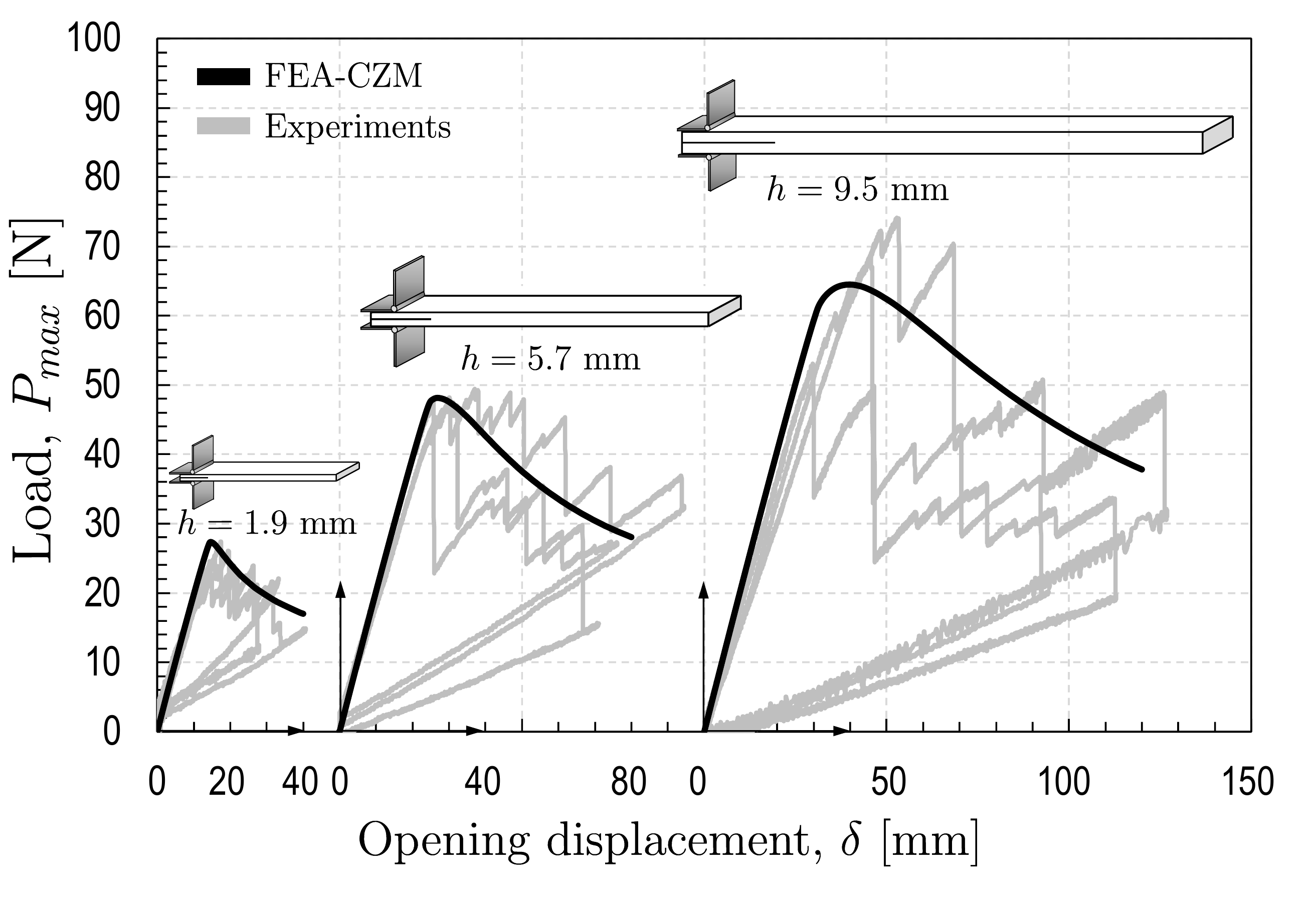}
\caption{Simulations by means of a Cohesive Zone Model (CZM) with a linear traction-separation law. The mode-I fracture energy used as input is estimated by means of Size Effect Law (SEL). }
\label{CZMI}
\end{figure}

The fracture energies used in the cohesive law were taken directly from SEL whereas the cohesive strength was calibrated to provide the best match of the experimental data, leading to a strength of $10$ MPa, for both the mode I and the mode II cohesive laws. As can be noted from Figure \ref{CZMI}, which shows a comparison between the experimental and predicted load displacement curves for the DCB specimens, the agreement with the test in terms of peak load is excellent. Further, the CZM was able to capture fairly well also the load-displacement evolution during crack propagation for all the specimen sizes. 

It should be mentioned that the jumps in crack propagation could not be captured by the model since the top and bottom layers of the FE mesh treated the composite as a homogenized continuum. Accordingly, phenomena such as crack deflection due to the waviness of the fabric were not considered explicitly. Also, it should be mentioned that an even better prediction of the crack propagation stage could have been achieved by using a different traction-separation law. In fact, recent results by Qiao and Salviato \cite{Yao2018b, Yao2018c} have shown that the cohesive behavior of thermoset polymers is best described by a bi-linear traction-separation law featuring a steep decrease of the cohesive stresses in  the first part of the curve followed by a milder reduction in the second (tail) branch of the cohesive law (this is similar to the cohesive softening law shape experimentally identified for concrete \cite{HooBaz14}).

\begin{figure}[H]
\center
\includegraphics[trim= 0cm 0cm 0cm 0cm,width=\textwidth]{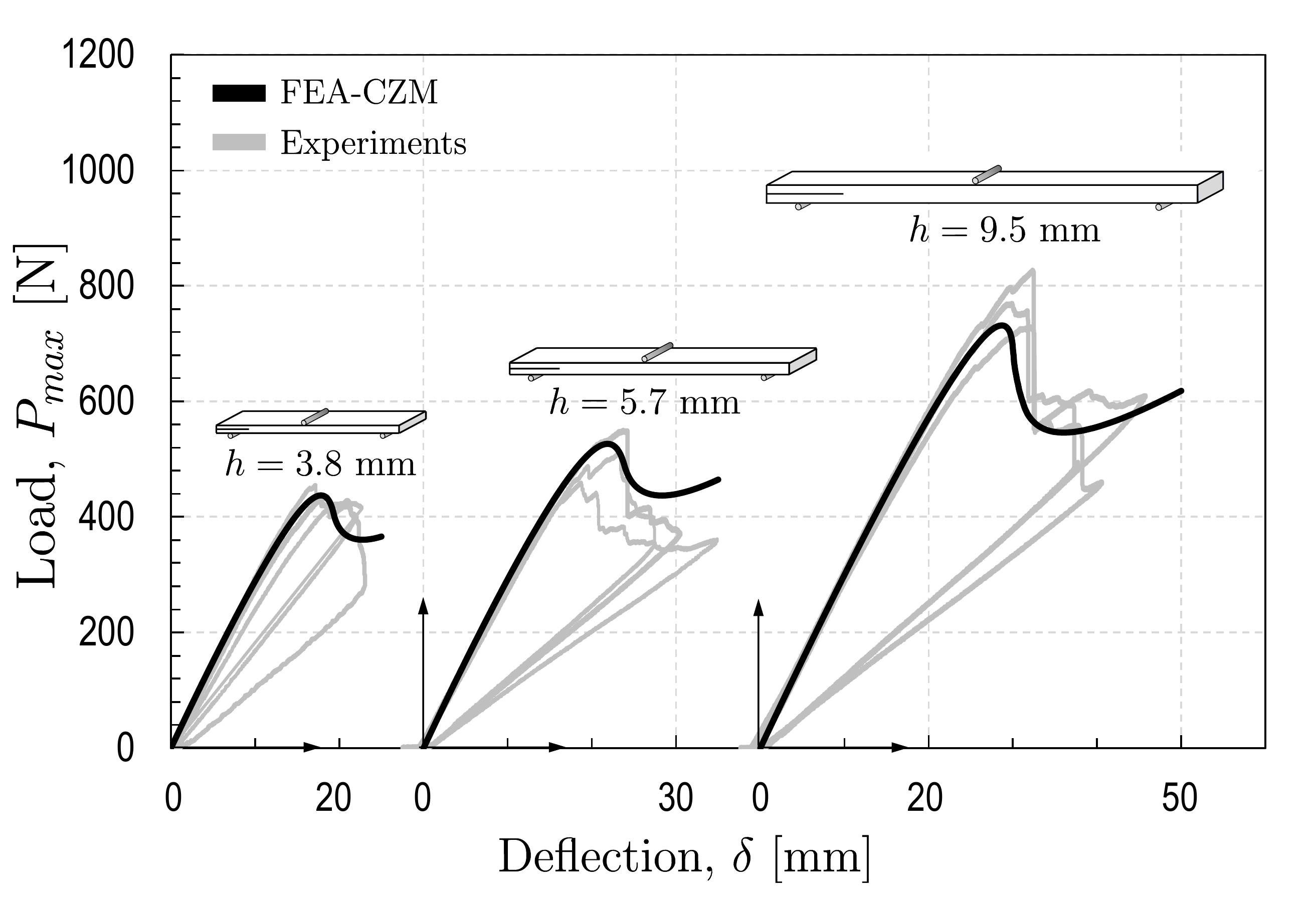}
\caption{Simulations by means of a Cohesive Zone Model (CZM) with a linear traction-separation law. The mode-II fracture energy used as input is estimated by means of Size Effect Law (SEL). }
\label{CZMII}
\end{figure}

Similar conclusions can be drawn for the ENF simulations. As shown in Figure \ref{CZMII}, the agreement between the Finite Element simulation and the test data in terms of peak loads is excellent. Moreover, also the post-peak behavior is captured very well with the predicted load-displacement curves becoming steeper and steeper with increasing specimen size. As can be noted from the experimental data, snap-back instability was occurring in the medium and large size specimens while the model predicted a steep but stable decrease of the load. Again, this may be due to the use of a linear, mode II cohesive law. The use of a bi-linear law with a steeper decrease of the shear stresses in the initial phase would lead to the prediction of a snap-back instability as well. 

The foregoing results are particularly important for the calibration of advanced computational models for composites. They confirm that the energy estimated by SEL, not the one calculated by LEFM, should be used as input in advanced computational models for composites. 

\section{Conclusions}

The present investigation of the mode I and II interlaminar fracturing behavior of laminated composites and the related size effect leads to the following conclusions:
\par
\vspace{0.2in}

1. The tests on geometrically scaled DCB and ENF specimens confirm a remarkable size effect for both mode I and II interlaminar fracturing. The analysis of the experimental data shows that, for the size range investigated in this work, the fracture scaling of the mode I interlaminar specimens is captured accurately by the Linear Elastic Fracture Mechanics (LEFM). However, this is not the case for mode II fracture, which exhibits a more complicated scaling. The double logarithmic plots of the nominal stress as a function of the characteristic size of the specimens show that the fracturing behavior evolves from pseudo-ductile to brittle with increasing sizes. For sufficiently large specimens, the size effect data tend to the classical $-1/2$ asymptotic slope predicted by LEFM. However, for smaller sizes, a significant deviation from the LEFM scaling is found, with the data exhibiting a milder scaling, which is a behavior associated with a more pronounced pseudo-ductility.
\vspace{0.2in}

2. Numerical simulations confirm that the more pronounced quasibrittleness of the ENF specimens is not associated to the frictional stresses acting on the crack surfaces. 

\vspace{0.2in}

3. The deviation from LEFM reported in the ENF experiments is related to the size of the Fracture Process Zone (FPZ). In mode I loading the damage/fracture zone close to the crack tip, characterized by significant non-linearity due to subcritical damaging, is generally very small compared to the specimen sizes investigated. This is in agreement with the inherent LEFM assumption of negligible nonlinear effects during the fracturing process. However, the damage mechanisms such as, e.g., matrix microcracking (Figure \ref{micro}a), crack deflection and plastic yielding occurring under mode II loading, lead to a significantly larger FPZ. For sufficiently small ENF specimens, the size of the highly non-linear FPZ is not negligible compared to the specimen characteristic size. and thus highly affects the fracturing behavior. This results into a significant deviation from the LEFM. 

\vspace{0.2in}

4. Capturing the correct scaling of the interlaminar fracturing behavior is of utmost importance for structural design. Further, it is a quintessential requirement to  measure correct material properties such as the mode I and II fracture energies. The analysis of the present results shows that the LEFM provides a relatively accurate description of the fracturing behavior and its scaling under mode I loading. In contrast, using LEFM to calculate the mode II fracture energy from the experiments leads to a size dependent $G_f^{(II)}$. In fact, the fracture energy obtained according to LEFM was 1.01 N/mm, 1.28 N/mm and 1.36 N/mm for the small, medium and large sizes, respectively. The reason for this discrepancy is that LEFM intrinsically lacks the  characteristic length and thus cannot capture the effects of the FPZ size.
\vspace{0.2in}

5. Following Ba\v{z}ant \cite{Baz84,Baz90,bazant1998_1}, an Equivalent Fracture Mechanics approach has been used to introduce a characteristic length, $c_f^{(i)}$, into the formulation. This length is related to the FPZ size and it is considered a material property, in addition to $G_f^{(i)}$. The resulting scaling equation, known as Ba\v{z}ant's Size Effect Law (SEL), depends not only on $G_f^{(i)}$ but also on the FPZ size. An excellent agreement with experimental data is shown, with SEL capturing the transition from quasi-ductile to brittle behavior as the size increases. The mode I fracture energy is found to be $0.56$ N/mm whereas the mode II fracture energy, defined  here as a material property independent of the specimen size, is $2.25$ N/mm. The equivalent FPZ lengths are found to be $c_f^{(I)}=$ 5.82 mm $c_f^{(II)}=$ 33.51 mm for modes I and II, respectively.
\vspace{0.2in}

6. The difference between the fracture energies predicted by LEFM and SEL depends on the FPZ size compared to the specimen size, with LEFM underestimating $G_f^{(i)}$ compared to SEL. For the DCB specimens investigated in this work, the difference between LEFM and SEL is 18$\%$, 4$\%$ and 2$\%$ for the small, medium and large sizes, respectively. For the ENF specimens, the LEFM predictions are about 55\%, 43\% and 40\% lower compared to the SEL for the small, medium and large sizes, respectively. The difference decreases with increasing specimen sizes and tends to zero for sufficiently large specimens, as the FPZ becomes negligible compared to the specimen size. 
\vspace{0.2in}

7. Finite Element simulations of the DCB and ENF tests by means of a cohesive zone model featuring a linear traction-separation law support the use of Size Effect Law (SEL) for the estimation of the mode I and II fracture energy. In fact, using the energy estimated by SEL as an input for the cohesive model, the agreement with the experimental load-displacement curves is excellent. This also confirms that the use of LEFM to calculate the fracture energy for cohesive zone models would lead to severe errors, especially in regards to the mode II cohesive law.
\vspace{0.2in}

8. The foregoing evidence shows that particular care should be devoted to the understanding of the scaling of the fracture behavior of laminated composites. In particular, the fracture tests carried out to characterize, e.g., the fracture energy, must guarantee objective results. Th size effect testing on geometrically scaled specimens is a simple and effective approach to provide objective results. The size effect method of measuring the mode I and II interlaminar fracture properties is easier to implement than other methods because only the peak load measurements are necessary: the post-peak behavior, crack tip displacement measurement and optical measurement of crack tip location are not needed. This is particularly advantageous for interlaminar fracture tests which are often affected by snap-back instability or discontinuous crack propagation, which make visual observations impractical and inaccurate.




\section*{Acknowledgments}
This material is based upon work supported by the Department of Energy under Cooperative Award Number DE-EE0005661 to the United States Automotive Materials Partnership, LLC and sub-award SP0020579 to Northwestern University.
The work was also partially supported under NSF grant No. CMMI-1435923 to Northwestern University. 

\clearpage
\listoftables
\listoffigures  
\clearpage

\section*{Figures and Tables}

\begin{table}[ht]
\centering  
\begin{tabular}{l c c} 
 \hline
  \rule{0pt}{4ex}
 Description & Symbol (units) & Measured value\\[1 ex]
 \hline
Fiber volume fraction & $V_f$ (-) & 0.54\\
Laminate thickness & $t$ (mm) & 1.9\\
In-plane modulus & $E$=$E_{1}$=$E_{2}$ (GPa)  & 53.5\\
In-plane shear modulus & $G$ = $G_{12}$ (GPa)  & 4.5 \\
In-plane Poisson ratio & $\nu=$$\nu_{12}$ = $\nu_{32}$ (-) & 0.055  \\
In-plane tensile strength in direction 1 and 2 & $F_{1t}$ = $F_{2t}$ (MPa) & 598 \\
\hline
\end{tabular}
\caption{\sf Properties of carbon twill 2x2/epoxy composite}
\label{T1}
\end{table}

\begin{table}[ht]
\centering  
\begin{tabular}{l c c c c}
\hline
\rule{0pt}{4ex}Size & Thickness, &Gauge length, &Crack length,  &Width, \\
&$h$&$L$&$a_0$&$W$\\[1 ex]
\hline
Small & $1.9$ & 100 & 40 & 25\\
Medium & $5.7$& 300 & 120 & 25\\
Large & $9.5$ & 500& 200& 25 \\
\hline
\multicolumn{5}{l}{Units: mm.}
\end{tabular}
\caption{\sf Geometrical specifications of the DCB specimens under study}
\label{T2}
\end{table}

\begin{table}[ht]
\centering  
\begin{tabular}{l c c c c}
\hline
\rule{0pt}{4ex}Size & Thickness, &Gauge length, &Crack length,  &Width, \\
&$h$&$L$&$a_0$&$W$\\[1 ex]
\hline
Small & $3.8$ & 200 & 50 & 25\\
Medium & $5.7$& 300 & 75 & 25\\
Large & $9.5$ & 500& 125& 25 \\
\hline
\multicolumn{5}{l}{Units: mm.}
\end{tabular}
\caption{\sf Geometrical specifications of the ENF specimens under study}
\label{T3}
\end{table}

\begin{table}[ht]
\centering  
\begin{tabular}{c c c c}
\hline
\rule{0pt}{4ex}Gauge length,  & Specimen thickness, &Max, load &Nominal strength\\
$L$ (mm)& $h$ (mm)& $P_{\mbox{max}}$ (N)& $\sigma_N$ (MPa)\\[1 ex]
\hline
\multirow{3}{*}{100}& \multirow{3}{*}{1.9}& 24.18 & 0.490 \\
 &                                          & 27.44 & 0.564 \\
 &                                          & 30.65 & 0.623 \\
\hline
\multirow{3}{*}{300}& \multirow{3}{*}{5.7}& 45.16 & 0.302 \\
&                                          & 44.03 & 0.297 \\
&                                       & 47.80 & 0.320 \\
\hline
\multirow{3}{*}{500}& \multirow{3}{*}{9.5} & 58.18 & 0.226 \\
&                                             & 70.86 & 0.245 \\
&                                            & 53.12 & 0.301 \\
\hline
\end{tabular}
\caption{\sf Results of tensile tests on Double Cantilever Beam (DCB) Specimens.}
\label{T4}
\end{table}

\begin{table}[ht]
\centering  
\begin{tabular}{c c c c}
\hline
\rule{0pt}{4ex}Gauge length,  & Specimen thickness, &Max, load &Nominal strength\\
$L$ (mm)& $h$ (mm)& $P_{\mbox{max}}$ (N)& $\sigma_N$ (MPa)\\[1 ex]
\hline
\multirow{3}{*}{200}& \multirow{3}{*}{3.8}& 450 & 4.74 \\
 &                                          & 400 & 4.21 \\
 &                                          & 390 & 4.11 \\
\hline
\multirow{3}{*}{300}& \multirow{3}{*}{5.7}& 512 & 3.59 \\
&                                          & 551 & 3.86 \\
&                                       & 460 & 3.24 \\
\hline
\multirow{3}{*}{500}& \multirow{3}{*}{9.5} & 828 & 3.49 \\
&                                             & 730 & 3.08 \\
&                                            & 769 & 3.28 \\
\hline
\end{tabular}
\caption{\sf Results of tensile tests on End Notch Flexure (ENF) Specimens.}
\label{T5}
\end{table}


\clearpage

\clearpage

\clearpage

\clearpage

\clearpage

\clearpage

\clearpage

\clearpage

\clearpage

\end{document}